# Melt Flow and Heat Transfer in Laser Drilling

Youqing Yang[a], Zhen Chen[b,a], Yuwen Zhang[c, 1]

[a] Department of Civil & Environmental Engineering, University of Missouri, Columbia, MO 65211, USA
[b] Faculty of Vehicle Engineering & Mechanics, State Key Laboratory of Structural Analysis for Industrial Equipment, Dalian University of Technology, Dalian 116024, China
[c] Department of Mechanical &Aerospace Engineering, University of Missouri, Columbia, MO 65211

**Abstract**
During the laser drilling process the recoil pressure drives melt flow and affects the heat transfer and material removal rate. To get a more realistic picture of the melt flow, a series of differential equations are formulated here that govern the process from pre-heating to melting and evaporation. In particular, the Navier-Stokes equation governing the melt flow is solved with the use of the boundary layer theory and integral methods. Heat conduction in solid is investigated by using the classical method with the corrections that reflect the change in boundary condition from the constant heat flux to Stefan condition. The dependence of saturation temperature on the vapor pressure is taken into account by using the Clausius-Clapeyron equation. Both constantly rising radial velocity profiles and rising-fall velocity profiles are considered. The proposed approach is compared with existing ones. In spite of the assumed varying velocity profiles, the proposed model predicts that the drilling hole profiles are very close to each other in a specific super alloy for given laser beam intensity and pulse duration.  The numerical results show that the effect of melt flow on material removal can be ignored in some cases. The findings obtained from the current work provide a better understanding of the effects of melt flow and vaporization on the laser drilling profile evolution, and could improve the solid material removal efficiency.

# Nomenclature

$a$, dimensionless curvature parameter of solid-liquid interface
$c_{pl}$, specific heat of the liquid [J kg$^{-1}$ K$^{-1}$]
$c_{ps}$, specific heat of the solid [J kg$^{-1}$ K$^{-1}$]
$g$, dimensionless melt layer thickness, normalized by laser beam radius
$h_{ls}$, latent heat of melting [J kg$^{-1}$]
$h_{lv}$, latent heat of vaporization[J kg$^{-1}$]
$H_{lv}$ , dimensionless latent heat of vaporization, $\frac{h_{lv}}{R_g T_{sat0}}$
$I_0$, laser intensity at the center [W m$^{-2}$]
$I'$, revised dimensionless laser intensity at the center for temperature calculation, $\frac{I_0}{p_0 h_{lv}}\sqrt{\frac{2\pi R_g T_{sat0}}{M}}$
$j_v$, is the molar flux of vaporization [kg s$^{-1}$]
$k_l$ thermal conductivity of liquid [Wm$^{-1}$ K$^{-1}$]

---

[1] Corresponding Author: zhangyu@missouri.edu

$k_s$, thermal conductivity of solid [Wm$^{-1}$ K$^{-1}$]

$k$, ratio of thermal conductivity of liquid over solid, $\frac{k_l}{k_s}$

$k'$, dimensionless coefficient, $\frac{k_l(T_{sat0}-T_m)}{RI_0}$

$M$, molar mass of the gas evaporated from the melt [kg mol$^{-1}$]

$N_i$, dimensionless laser intensity of laser beam, $\frac{RI_0 c_{pl}}{h_{lv} k_l}$

$N_\alpha$, thermal diffusivity ratio, $\frac{\alpha_s}{\alpha_l}$

$N_c$, specific heat ration, $\frac{c_{ps}}{c_{pl}}$

$p_c$, vapor pressure at the center of laser beam

$Pr$, Prattle Number, $\frac{\mu}{\rho \alpha_l}$

$p_0$, vapor pressure at saturation temperature, 1.013×10$^5$ Pa

$R$, Gauss radius of laser beam, defined as the intensity reduced to *1/e* of that at the central point [m]

$R_g$ gas constant of the metal vapor [J kg$^{-1}$ K$^{-1}$]

$R_h$, latent heat ratio, $\frac{h_{lv}}{h_{ls}}$

$Sc$, subcooling parameter, $\frac{c_{ps}(T_m-T_i)}{h_{ls}}$

$Ste$, Stefan number, $\frac{c_{pl}(T_{sat0}-T_m)}{h_{ls}}$

$T_i$, initial temperature of the solid [K]

$T_m$, melting temperature at solid-liquid interface [K]

$T_{sat0}$, saturation temperature at pressure $p_0$, [K]

$T_{sat}$, saturation temperature at pressure $p$, [K]

$t_p$, pulse on time, [s]

$U$, dimensionless radial velocity in the free flow

$V$, dimensionless vertical velocity in the free flow

$V_v$, vapor velocity at the melt surface

$u$, tangential velocity [m s$^{-1}$]

$v$, normal velocity [m s$^{-1}$]

$\alpha_l$, thermal diffusivity of melt, $\frac{k_l}{\rho c_{pl}}$

$\theta_s$, dimensionless temperature in solid, $\frac{T_s-T_m}{T_{sat0}-T_m}$

$\theta_l$, dimensionless temperature in liquid, $\frac{T_l-T_m}{T_{sat0}-T_m}$

$\theta_i$, dimensionless initial temperature of solid, $\frac{T_i-T_m}{T_{sat0}-T_m}$

$\theta_m$, ratio of melting temperature over saturation temperature, $\frac{T_m}{T_{sat0}}$

$\theta_{sat}$, dimensionless temperature at the melt-vapor interface, $\frac{T_{sat}-T_m}{T_{sat0}-T_m}$

$\tau$, dimensionless time, $\frac{t\alpha_l}{R^2}$

# 1. Introduction

There are two major mechanisms that control the material removal in the process of laser drilling: (1) melt evaporation, and (2) melt expulsion by the vaporization-induced recoil force. It is generally thought that melt removal dominates if an assisting gas is applied on the melt surface when the melt surface temperature does not significantly exceed the melting point and the evaporation rate is low enough not to produce a noticeable recoil pressure.  In the cases where there is no assisting gas involved, melt expulsion varies with the recoil pressure, which is highly dependent on the surface temperature. At a high surface temperature, the melt removal due to evaporation may exceed that by the hydrodynamic mechanism due to the recoil force. In an early simulation, a significant portion of the absorbed laser intensity was found to be taken away by the melt flow from the heat interaction zone [1]. Due to the difficulty in directly measuring the interface geometry and the temperature and recoil pressure at the melt-vapor interface, the portion of the melt removed by the recoil-force-driven flow cannot be quantitatively determined with existing experimental capabilities.

Considerable research has been carried out to develop a theoretical model for predicting the laser drilling response. Assuming a constant laser beam intensity profile, von Allmen analyzed the drilling velocity and drilling efficiency by using a one-dimensional (1-D) transient gas dynamic model [2]. Chan and Mazumder [3] developed a 1-D steady state model to incorporate liquid expulsion, but the 1-D assumption is not suited for the hole drilling with high aspect ratio and the drilling process is transient. Kar and Mazumder [4] extended the model to two-dimensional (2-D) cases in which melt expulsion was not considered. Armon et al. formulated a 1-D metal drilling problem based on the enthalpy balance method and solved the problem by using the Crank-Nicholson method [5]. They also conducted an experimental investigation on metal drilling with a $CO_2$ laser beam and analyzed the experimental results by using their theoretical model [6]. A more rigorous treatment of melt expulsion was presented by Ganesh et al. [7], which employed a 2-D transient generalized model and incorporated conduction, convection and phase change heat transfer during laser drilling; this model, however, is computationally demanding. Zhang and Faghri developed an analytical model to study the effect of solid conduction on the material removal rate and phase change at interfaces [8]. In this model, the melt flow effect on heat transfer is neglected. Zhang et al. developed a 2-D transient model, in which a Knudsen layer was considered at the melt-vapor front without including the melt flow effect [9]. Pastras et al. analyzed the material removal efficiency by assuming linear temperature profiles in solid, liquid and vapor [10], with an implicit assumption that the melt flow does not cause any disturbance on temperature gradient.

The melt flow effect has been considered in some existing models. For example, the model developed by Semak and Matsunawa [1] and a later version adapted by Low et al. [11] to include the melt flow effect with an assisting gas on laser drilling are both steady-state based on conservation of mass and energy. Semak and Matsunawa attempted to evaluate the effect of recoil pressure during the melt ejection process, and their model is based on the assumption of a free flow layer of melt under the laser beam of hat-top shaped intensity profile [1]. They also considered the temperature-dependent pressure (but not Clausius-Clapeyron equation). Ng et al. developed a model of laser drilling incorporating the effect of using oxygen as an assisting gas. They assumed that the melt front propagates with an averaged velocity and the averaged melt thickness is determined via dividing the thermal diffusivity of the melt by the averaged propagating velocity [12]. Zeng et al. developed a 2-D analytical model for optical trepanning assuming that vaporization rate is negligible [13]. Collins and Gremaud developed a simple 1-D model by cross-section averaging while neglecting the contribution of the radial flow velocity component [14]. It is worthy to note that the melt flow models developed in [1, 2, 12] and the latest simulation by Semak and Miller [15] all assume a hat-top-shaped intensity profile. The assumption about the laser beam intensity profile directly affects the conclusion about the melt flow [16-18]. Using the hat-top profile, the melt surface temperature could be assumed to be constant, though a rapid change occurs at the margin of the melt. If the melt flow is further assumed to be free of shear traction, the recoil pressure can also be assumed to be constant, which leads to an overestimate of the role of melt expulsion. Hence, the melt flow effect on laser drilling should be reevaluated based on a more realistic model.

A more realistic model should consider vaporization based on the real physics involved. It is known that vaporization occurs at any temperature above the melting point, and that the recoil pressure is highly dependent on the melt surface temperature. However, some previous models assumed a Stefan condition at the melt-vapor interface [7], while some others took the boiling point for the liquid-vapor transition [19]. Solana et al. assumed the recoil pressure to be of the Gaussian form [20]. Li et al. assumed that the liquid-vapor transition takes place over a certain temperature range [21].

How to simulate heat conduction more accurately is also important to better predict the real physics. Heat conduction in solid is a classical problem, but the heat conduction in laser drilling involves a change in boundary conditions, which has led to different approaches by different investigators. Earlier researchers assumed a constant melt layer thickness and a constant melting rate, and consequently developed a steady state heat conduction model [22]. Modest derived a transient heat conduction model by assuming that the phase change from solid to vapor occurs in a single step [23]. By assuming a parabolic temperature profile and applying integration, the partial differential equation was transformed into an ordinary differential equation, which was later applied for an integral solution by Zhang and Faghri [8]. Shen et al. also derived a transient heat conduction model by assuming a temperature profile of exponential function [24]. Ho and Lu developed a transient heat conduction model by adding a heat source term in the solid to represent the energy flux from the laser beam [25]. Shidfar et al. developed a transient heat conduction model by assuming that the solid being heated is initially at the melting point [26]. These models have been frequently cited in the laser simulation research community, and helpful in understanding the physics with different degrees of success, but they have all had certain limitations due to the assumptions made. Laser drilling in a solid is a transient heat conduction process with the thermal energy emanating from laser beams, and hence, classical methods [27] are available for us to develop a theoretical model that might better predict the process with few assumptions.

In this work, we aim at developing a theoretical model with the melt flow effect being explicitly included. Once the solution about the melt flow velocity becomes available, the melt flow effect on the heat transfer can be evaluated with confidence. The previous assumptions such as ignoring melt flow can thus be evaluated using the proposed method. We also try to keep the model as simple as possible to reduce computational expense. Unlike previous studies, the melt is assumed to behave like a Newtonian fluid, and the non-slip boundary condition is applied along the solid-liquid interface. With a Gaussian intensity profile being assumed for the laser beam, the vapor temperature, recoil pressure and initial melting time all vary in the radial direction. Heat absorbed in the solid and transferred by phase change and melt flow are all taken into account. Both boundary layer formulation and integral forms of momentum equation and energy equation are developed. Finally, a set of numerical experiments are performed for a special super alloy with the proposed approach, and compared with the representative experimental data and solutions predicted by other models.

## 2. Interface energy balance and governing equations

Figure 1 shows the coordinate system used for formulating the equations, in which $r$ denotes the radial direction, and $Z$ the upward direction pointing to the melt from the solid, with origin at the solid-melt interface. Note that we use the lower-case letter $z$ to mark the change of interface between phases, and $z$ originates at the initial solid surface and always points toward the solid. The solid-melt interface is curved in nature, but for the first order of approximation, we simply ignore the difference between tangent direction and radial direction in current study on melt flow, similar to that by Ganesh et al. [7]. The local coordinate $n$ is defined to originate at the solid-melt interface and point to the solid in the normal direction.

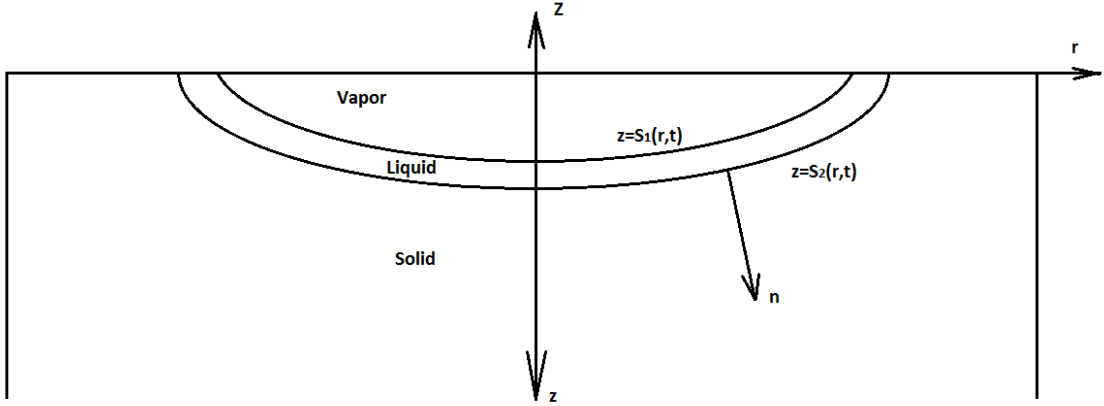

Fig. 1. Physical model of laser drilling and the coordinate systems used.

## 2.1 Melt convection

In the melt layer, the vertical velocity is much smaller than the lateral component, which enables us to simplify the governing equations considerably. We also ignore the friction produced heat, surface tension and gravitational force. Similar to the assumptions in the jet impingement study by Kendoush [28], the continuity equation, momentum equation and energy equation can be written as:

$$\frac{\partial u}{\partial r} + \frac{\partial v}{\partial Z} + \frac{u}{r} = 0 \tag{1}$$

$$u\frac{\partial u}{\partial r} + v\frac{\partial u}{\partial Z} = \frac{\mu}{\rho}\left(\frac{\partial^2 u}{\partial Z^2} + \frac{1}{r}\frac{\partial u}{\partial r} - \frac{u}{r^2}\right) - \frac{1}{\rho}\frac{\partial p}{\partial r} \tag{2}$$

$$\frac{\partial T_l}{\partial t} + u\frac{\partial T_l}{\partial r} + v\frac{\partial T_l}{\partial Z} = \alpha_l \frac{\partial^2 T_l}{\partial Z^2} \tag{3}$$

where $u$ and $v$ are the radial, and vertical velocity component, respectively. $\rho$ represents the density of the melt and $\mu$ the viscosity. $T_l$ is the temperature in the liquid and $\alpha_l$ is the thermal diffusivity. The lower case $t$ denotes time and $p$ represents pressure. For the convenience of the description of boundary layer theory, $U$ and $V$ stand for free stream flow velocity outside of the boundary layer, and the radial component $U$ does not change in the Z-direction. From Bernoulli's principle, we have the following equation for the free stream,

$$-\frac{1}{\rho}\frac{\partial p}{\partial r} = U\frac{dU}{dr} \tag{4}$$

Considering that the melt layer is much thinner compared with its lateral dimension, we assume that the pressure variation in the thickness direction is ignorable, thus, the second momentum conservation equation becomes $\frac{\partial p}{\partial Z} = 0$, so pressure can be treated as a function of radius only. Defining $\xi = \frac{r}{R}, \zeta = \frac{Z}{R}, \tau = \frac{t\alpha_l}{R^2}, \alpha_l = \frac{k_l}{\rho c_{pl}}, u^* = \frac{uR}{\alpha_l}, v^* = \frac{vR}{\alpha_l}, U^* = \frac{UR}{\alpha_l}, p^* = \frac{pR^2}{\rho \alpha_l^2},$ $\theta_l = \frac{T_l - T_m}{T_{sato} - T_m},$ we rewrite the above equations in a non-dimensional form

$$\frac{\partial u^*}{\partial \xi} + \frac{\partial v^*}{\partial \zeta} + \frac{u^*}{\xi} = 0 \tag{1a}$$

$$u^* \frac{\partial u^*}{\partial \xi} + v^* \frac{\partial u^*}{\partial \zeta} = Pr \left( \frac{\partial^2 u^*}{\partial \zeta^2} + \frac{1}{\xi} \frac{\partial u^*}{\partial \xi} - \frac{u^*}{\xi^2} \right) - \frac{dp^*}{d\xi} \quad (2a)$$

$$\frac{\partial \theta_l}{\partial \tau} + u^* \frac{\partial \theta_l}{\partial \xi} + v^* \frac{\partial \theta_l}{\partial \zeta} = \frac{\partial^2 \theta_l}{\partial \zeta^2} \quad (3a)$$

$$-\frac{\partial p^*}{\partial \xi} = U^* \frac{dU^*}{d\xi} \quad (4a)$$

For convenience, we simply remove the asterisk from each variable now and hereafter. The $-\frac{dp}{d\xi}$ is also replaced with $U \frac{dU}{d\xi}$. Now we have

$$\frac{\partial u}{\partial \xi} + \frac{\partial v}{\partial \zeta} + \frac{u}{\xi} = 0 \quad (1b)$$

$$u \frac{\partial u}{\partial \xi} + v \frac{\partial u}{\partial \zeta} = Pr \left( \frac{\partial^2 u}{\partial \zeta^2} + \frac{1}{\xi} \frac{\partial u}{\partial \xi} - \frac{u}{\xi^2} \right) + U \frac{dU}{d\xi} \quad (2b)$$

$$\frac{\partial \theta_l}{\partial \tau} + u \frac{\partial \theta_l}{\partial \xi} + v \frac{\partial \theta_l}{\partial \zeta} = \frac{\partial^2 \theta_l}{\partial \zeta^2} \quad (3b)$$

Assuming there exists a boundary layer of melt of thickness $\delta$, we have the boundary conditions as follows

$$v = 0 \ at \ \zeta = 0 \quad (5)$$

$$u = U \ at \ \zeta = \delta \quad (6)$$

$$\frac{\partial u}{\partial \zeta} = 0 \ at \ \zeta = \delta \quad (7)$$

$$u = 0 \ at \ \zeta = 0 \quad (8)$$

$$u = 0 \ at \ \xi = 0 \quad (9)$$

$$\theta_l = 0 \ at \ \zeta = 0 \quad (10)$$

$$\theta_l = \theta_{sat} \ at \ \zeta = g(\xi, \tau) \quad (11)$$

## 2.2 Pressure-dependent saturation temperature

In the following sessions, $z = s_1(r,t)$ and $z = s_2(r,t)$ are the vapor-melt interface and melt-solid interface respectively. At $z = s_1(r,t)$, energy conservation requires that the sum of vapor kinetic energy, latent heat due to vaporization, and sensible heat into the melt be equal to the input energy from laser beam, i.e.

$$\frac{1}{2} j_v M V_v^2 + h_{lv} \rho \frac{\partial s_1}{\partial t} + k_l \frac{\partial T_l}{\partial z} \Big|_{z=s_1} = I \quad (12)$$

where $M$ is the molar mass of the gas, $j_v$ is the molar flux of evaporation, $h_{lv}$ denotes the latent heat of vaporization, and $k_l$ represents the heat conductivity of the liquid. $I$ is the laser beam intensity. Eq. (12) actually is the equilibrium of energy per unit time and per unit area. Assuming the temperature is continuous at the melt-vapor interface, the vapor velocity at the interface is

$$V_v = \sqrt{\gamma R_g T_{sat}/M} \tag{13}$$

where $\gamma = \frac{c_p}{c_v} = \frac{5}{3}$ for a monoatomic ideal gas [29]. From Eq. (12), one gets

$$\rho \frac{\partial s_1}{\partial t} = \frac{I - \frac{1}{2} j_v M V_v^2 - k_l \frac{\partial T_l}{\partial z}|_{z=s_1}}{h_{lv}} \tag{14}$$

The vapor pressure is related to the temperature by the Clausius-Clapeyron equation:

$$p(T_{sat}) = p_0 \exp\left[\frac{h_{lv}}{R_g}\left(\frac{1}{T_{sat0}} - \frac{1}{T_{sat}}\right)\right] \tag{15}$$

where $T_{sat}$ denotes the vapor temperature and $T_{sat0}$ is the saturation temperature at pressure $p_0$. $R_g$ is the gas constant. The vapor leaving the melt surface can also be counted by using the product of molar flux and molar mass.

$$\rho \frac{\partial s_1}{\partial t} = j_v M \tag{16}$$

Bellantone and Ganesh [29] obtained the following relation:

$$j_v = \frac{p}{\sqrt{2\pi M R_g T_{sat}}} \tag{17}$$

Substituting Eq. (16) into Eq. (14), one gets

$$j_v = \frac{I - k_l \frac{\partial T_l}{\partial z}|_{z=s_1}}{h_{lv} M + \frac{1}{2}\gamma R_g T_{sat}} \tag{18}$$

From the right sides of Eq. (18) and Eq. (17), one has

$$p = \sqrt{\frac{2\pi R_g T_{sat}}{M}} \frac{I - k_l \frac{\partial T_l}{\partial z}|_{z=s_1}}{h_{lv} + \frac{\gamma R_g}{2M} T_{sat}} \tag{19}$$

Substituting Eq. (15) into Eq. (19), one gets

$$p_0 \exp\left[\frac{h_{lv}}{R_g}\left(\frac{1}{T_{sat0}} - \frac{1}{T_{sat}}\right)\right] = \sqrt{\frac{2\pi R_g T_{sat}}{M}} \frac{I - k_l \frac{\partial T_l}{\partial z}|_{z=s_1}}{h_{lv} + \frac{\gamma R_g}{2M} T_{sat}} \tag{20}$$

The laser beam intensity takes the Gaussian form as follows

$$I(r,t) = I_0 \exp\left(-\frac{r^2}{R^2}\right) = I_0 \exp(-\xi^2), 0 < t < t_p \tag{21}$$

Using dimensionless parameters $k' = \frac{k_l(T_{sat0} - T_m)}{R I_0}$ and $I' = \frac{I_0}{p_0 h_{lv}} \sqrt{\frac{2\pi R_g T_{sat0}}{M}}$, $\Lambda = \frac{\gamma R_g T_{sat0}}{2M h_{lv}}$, $\theta_{sat} = \frac{T_{sat} - T_m}{T_{sat0} - T_m}$, $\theta_m = \frac{T_m}{T_{sat0}}$, we simplify Eq. (20) as

$$\{1 + \Lambda[\theta_{sat}(1-\theta_m) + \theta_m]\}exp\left[H_{lv}\left(1 - \frac{1}{\theta_{sat}(1-\theta_m)+\theta_m}\right)\right] =$$
$$I'\sqrt{\theta_{sat}(1-\theta_m) + \theta_m}\left[exp(-\xi^2) - k'\frac{\partial \theta_l}{\partial \zeta}|_{\zeta=S_1}\right] \tag{22}$$

## 2.3 Energy balance at solid-melt interface

Both the energy for heating up the solid and melting are from the liquid; thus, the energy balance at the solid-liquid interface can be written as follows:

$$\rho h_{ls}\frac{\partial s_2}{\partial t} = k_s\frac{\partial T_s}{\partial z}|_{z=s_2} - k_l\frac{\partial T_l}{\partial z}|_{z=s_2} \tag{23}$$

where $h_{ls}$ denotes the latent heat of melting and $k_s$ and $k_l$ represent the thermal conductivity of the solid and the fluid, respectively. In this equation, it is assumed that the density of the solid is the same as that of the fluid. The dimensionless form is

$$\frac{\partial S_2}{\partial \tau} = Ste N_\alpha N_c\left(-\frac{\partial \theta_s}{\partial \zeta}|_{\zeta=0} + k\frac{\partial \theta_l}{\partial \zeta}|_{\zeta=0}\right) \tag{24}$$

where $Ste = \frac{c_{pl}(T_{sat0}-T_m)}{h_{ls}}$ is the Stefan number; $N_\alpha = \frac{\alpha_s}{\alpha_l}$ is the thermal diffusivity ratio; and $N_c = \frac{c_{ps}}{c_{pl}}$ is the specific heat ration; $k = \frac{k_l}{k_s}$ represents the ratio of liquid thermal conductivity over solid.

Eq. (14) is the energy balance at the melt-vapor interface. From Eqs. (13) and (17), the vapor kinetic energy $\frac{1}{2}j_v M V_v^2 = \frac{p\gamma}{2}\frac{\sqrt{R_g T_{sat}}}{\sqrt{2\pi M}} = \frac{p\sqrt{\gamma\Lambda h_{lv}}}{2\sqrt{\pi}}\sqrt{\frac{T_{sat}}{T_{sat0}}}$, and let $\Psi = \frac{\sqrt{\gamma\Lambda}}{2\sqrt{\pi}}\frac{\alpha_l}{R\sqrt{h_{lv}}}$; hence, the dimensionless form of (14) is

$$\frac{\partial S_1}{\partial \tau} = N_i \exp(-\xi^2) - \Psi p\sqrt{\theta_{sat}(1-\theta_m) + \theta_m} - \frac{h_{ls}}{h_{lv}}Ste\frac{\partial \theta_l}{\partial \zeta}|_{\zeta=g} \tag{25}$$

Note that the pressure $p$ in the Eq. (25) is dimensionless whereas the one in the (15), (17), (19) is dimensional.

## 2.4 Heat conduction in solid

To analyze the heat conduction in the solid, we set up a local one-dimensional coordinate system $n$ that originates at the melt-solid interface and points to the solid in the normal direction (see Fig. 1). Before the surface temperature reaches the melting point, the solid surface receives constant heat flux, and the solution is [27]:

$$\theta_s(\xi, n, \tau) = \theta_i + \frac{k}{k'}exp(-\xi^2)\sqrt{N_\alpha \tau}\left[\frac{1}{\sqrt{\pi}}exp(-\eta^2) - \eta erfc(\eta)\right] \tag{26}$$

where $\theta_s = \frac{T_s - T_m}{T_{sat0}-T_m}$ is the dimensionless temperature in the solid whereas $T_s$ corresponds to the dimensional value; $\theta_i = \frac{T_i - T_m}{T_{sat0}-T_m}$ is the dimensionless initial temperature of the solid, and $T_i$ is the dimensional value, $T_m$ is the melting point, and $k' = \frac{k_l(T_{sat0}-T_m)}{RI_0}$, and $\eta = \frac{n}{2\sqrt{N_\alpha \tau}}$, which is referred to as the similarity variable.

It is obvious that the highest temperature is at the surface. When $T_s = T_m$, $\theta_s = 0$, phase change begins at the surface. The melting starting time can thus be obtained from Eq. (26)

$$\tau_m = \frac{\pi}{N_\alpha}\left[\frac{k'\theta_i}{k}\exp(\xi^2)\right]^2 \tag{27}$$

At the moment $\tau = \tau_m$, the following temperature profile is established.

$$\theta_s(\xi, n, \tau_m) = \theta_i - \theta_i\left[\exp(-\eta_m^2) - \sqrt{\pi}\eta_m \mathrm{erfc}(\eta_m)\right] \tag{28}$$

where $\eta_m = \frac{n}{2\sqrt{N_\alpha \tau_m}}$.

The exact solution for time period after melting starts were derived in this study as well as the geometrical correction for the temperature gradient (see Appendix). In short, it consists of three parts: (1) the solution for Stefan boundary condition; (2) the correction for the difference between the temperature profile due to heating at the first stage and that for Stefan condition; and (3) further correction to assure the constant temperature condition at the solid-melt interface.

$\frac{\partial \theta_s}{\partial \zeta}$ in Eq. (24) is related to $\frac{\partial \theta_s}{\partial n}|_{n=0}$ by considering the heat energy fluency crossing the melt surface, which gives

$$\frac{\partial \theta_s}{\partial \zeta}\bigg|_{n=0} = -C_c\left[\frac{\partial \theta_s}{\partial n}(0,\tau)\sqrt{1 + \left(\frac{\partial S_2}{\partial \xi}\right)^2} + \frac{\theta_i}{N_\alpha}\frac{\partial S_2}{\partial \tau}\right] \tag{29}$$

where $C_c$ is the curvature correction coefficient as defined in the Appendix. Substituting (29) and the definition of $V_n$ into (24) leads to

$$\frac{1}{N_\alpha}\left(\frac{1}{SteN_c} - C_c\theta_i\right)\frac{\partial S_2}{\partial \tau} = C_c\frac{\partial \theta_s}{\partial n}(0,\tau)\sqrt{1 + \left(\frac{\partial S_2}{\partial \xi}\right)^2} + k\frac{\partial \theta_l}{\partial \zeta}\bigg|_{\zeta=0} \tag{30}$$

## 3. Solution of the melt flow

Solving the melt flow field is the key to more accurately evaluate the role of melt expulsion. Since the pressure that drives the flow is dependent to the temperature, equations (24) and (30) cannot be solved alone. The temperature gradient is influenced by the melt flow as shown in the equation of energy equation (3b). Solving the flow field becomes the key point in seeking a more accurate temperature solution, which makes this research unique by comparing with previous researches. In this section, we focus on solving the momentum equation. Successful applications for this purpose can be found in natural and forced convection near fluid-solid interface by using boundary-layer theory [30]. Our first idea was to seek the solution by applying boundary layer theory.

## 3.1 Boundary layer formulation

Integrating Eq. (1b) with respect to $\zeta$ from 0 to $\delta$, one gets the velocity at the top of the boundary layer:

$$v_\delta = -\int_0^\delta \left(\frac{\partial u}{\partial \xi} + \frac{u}{\xi}\right) d\zeta \qquad (31)$$

Note that

$$\int_0^\delta v \frac{\partial u}{\partial \zeta} d\zeta = -u(\xi,\delta) \int_0^\delta \left(\frac{\partial u}{\partial \xi} + \frac{u}{\xi}\right) d\zeta + \int_0^\delta u \left(\frac{\partial u}{\partial \xi} + \frac{u}{\xi}\right) d\zeta \qquad (32)$$

Integrating the momentum equation (2b) with respect to $\zeta$ from 0 to $\delta$, results in

$$\int_0^\delta u \frac{\partial u}{\partial \xi} d\zeta - u(\xi,\delta) \int_0^\delta \left(\frac{\partial u}{\partial \xi} + \frac{u}{\xi}\right) d\zeta + \int_0^\delta u \left(\frac{\partial u}{\partial \xi} + \frac{u}{\xi}\right) d\zeta = \int_0^\delta Pr \left[\frac{\partial^2 u}{\partial \zeta^2} + \frac{1}{\xi}\frac{\partial u}{\partial \xi} - \frac{u}{\xi^2}\right] d\zeta + \int_0^\delta U \frac{dU}{d\xi} d\zeta \qquad (33)$$

From $-\frac{dp}{d\xi} = U\frac{dU}{d\xi}$, we have $\frac{1}{2}U^2 + p = Constant$. At the center of laser beam, $\xi = 0, U = 0, p = p_c$, where $p_c$ is the pressure at the beam center, thus, one gets

$$U = \sqrt{2(p_c - p(\xi))} \qquad (34)$$

Beneath the free flow, the radial velocity is assumed to be in the form of

$$u = U\left(2\frac{\zeta}{\delta} - \frac{\zeta^2}{\delta^2}\right) \qquad (35)$$

which automatically meets boundary conditions of (6) to (9).

Substituting Eq. (35) and the derivatives into (33), one gets

$$\left(Pr - \frac{16}{5}\xi U\right)\frac{d\delta}{d\xi} = -Pr\left(\frac{3\xi}{\delta} - \frac{\delta}{U}\frac{dU}{d\xi} + \frac{\delta}{\xi}\right) + \frac{3}{4}\xi\delta\frac{dU}{d\xi} + \frac{1}{8}\delta U \qquad (36)$$

This differential equation can be solved using the Runge-Kutta method [31] with a certain boundary condition. With zero free flow velocity at the center, it seems that any value of the boundary layer thickness will work. This uncertainty forces our attention to the second node. By assuming $\frac{d\delta}{d\xi} = 0$ at the beam center which reflects the axisymmetric condition and approximating $\frac{dU}{d\xi}$ with $\frac{U}{\xi}$, we get $\delta = \sqrt{\frac{24\xi Pr}{7U}}$ from Eq. (36). Note this formula holds approximately only at the second node. Using this boundary value, the succeeding values will be solved with Runge-Kutta method being applied to Eq. (36).

The boundary layer flow velocity profile may also be assumed to be a cubic function as follows

$$u = U\left(\frac{3}{2}\frac{\zeta}{\delta} - \frac{1}{2}\frac{\zeta^3}{\delta^3}\right), \qquad (37)$$

which automatically meets the boundary conditions in (6) to (9).

Integrating the continuity equation, one gets

$$v = -\int_0^\zeta \left(\frac{\partial u}{\partial \xi} + \frac{u}{\xi}\right) d\zeta = -\frac{\delta}{8}\left(\frac{dU}{d\xi} + \frac{U}{\xi}\right)\left[6\left(\frac{\zeta}{\delta}\right)^2 - \left(\frac{\zeta}{\delta}\right)^4\right] + \frac{3}{8}U\frac{d\delta}{d\xi}\left[2\left(\frac{\zeta}{\delta}\right)^2 - \left(\frac{\zeta}{\delta}\right)^4\right]$$

(38)

Substituting the above and Eq. (37) into Eq. (33) one gets

$$(105Pr - 39\xi U)\frac{d\delta}{d\xi} = -35Pr\left(\frac{12\xi}{\delta} - 5\frac{\delta}{U}\frac{dU}{d\xi} + 5\frac{\delta}{\xi}\right) + 39\delta U + 183\xi\delta\frac{dU}{d\xi} \,,$$ (39)

which can be solved by using the boundary condition $\delta = \sqrt{\frac{70Pr\xi}{37U}}$ at the second node similar to the reasoning discussed in the previous paragraph for Eq. (36).

The free flow $U$ could lead to the melt layer thickness change. The mass conservation requires

$$\frac{\partial U}{\partial \xi} + \frac{\partial V}{\partial \zeta} + \frac{U}{\xi} = 0$$ (40)

Integrated Eq. (40) over $\zeta$ from $\zeta = \delta$ to $\zeta = g$, one has

$$V = -\left(\frac{\partial U}{\partial \xi} + \frac{U}{\xi}\right)(g - \delta)$$ (41)

The vertical velocity at the surface of the boundary layer is

$$v_\delta = -\int_0^\delta \left(\frac{\partial u}{\partial \xi} + \frac{u}{\xi}\right) d\zeta$$ (42)

For the profile represented in Eq. (35), $v_\delta = -\frac{2}{3}\delta\left(\frac{\partial U}{\partial \xi} + \frac{U}{\xi}\right)$

For the profile represented in Eq. (37), $v_\delta = -\frac{5}{8}\delta\left(\frac{\partial U}{\partial \xi} + \frac{U}{\xi}\right)$

The melt layer thickness varies due to melting rate, evaporating rate, and the flow-related changes as shown below:

$$\frac{\partial g}{\partial \tau} = \frac{\partial S_2}{\partial \tau} - \frac{\partial S_1}{\partial \tau} + V + v_\delta$$ (43)

which can be used to update the melt layer thickness, i.e.

$$g = g_0 + (\frac{\partial S_2}{\partial \tau} - \frac{\partial S_1}{\partial \tau} + V + v_\delta)\Delta\tau$$ (44)

where $g_0$ denotes the melt thickness at the previous time step and $g$ at current step. $\Delta\tau$ is the time interval.

Note that $\frac{\partial S_1}{\partial \tau}$ represents the melt surface change due to vaporization only. Considering vertical displacement due to flow, we get the net melt-vapor interface

$$S_{1net} = S_2 - g \qquad (45)$$

The space between the initial solid surface and the net melt surface will be counted as the material removed by the laser beam because the melt will be recast once the heating process ends.

## 3.2 Integration method formulation

Near the laser beam center, the boundary layer is thin. Far from the beam center, the boundary layer increases and the free flow layer may disappear. In this case, because boundary conditions (6) and (7) no longer remain valid, we return to Eq. (2a). We now seek an integral solution. Eq. (33) is rewritten as

$$\int_0^g u \frac{\partial u}{\partial \xi} d\zeta - u(\xi, g) \int_0^g \left( \frac{\partial u}{\partial \xi} + \frac{u}{\xi} \right) d\zeta + \int_0^g u \left( \frac{\partial u}{\partial \xi} + \frac{u}{\xi} \right) d\zeta = \int_0^g Pr \left( \frac{\partial^2 u}{\partial \zeta^2} + \frac{1}{\xi} \frac{\partial u}{\partial \xi} - \frac{u}{\xi^2} \right) d\zeta - g \frac{dp}{d\xi} \qquad (46)$$

The radial velocity must have experienced a rising course since it is zero at the laser beam center. However, the radial velocity profile cannot be $u = A\xi\zeta$, because this function will lead to $\frac{dp}{d\xi} = 0$, which means the recoil pressure is constant in the melt zone. We cannot assume a profile like $u = A\zeta\xi^2$ or any power index larger than 1 because a free flow cannot be that way with respect to the distance. If the radial velocity is $u = A\zeta\xi^{1/m}$ where $m > 1$ and $A$ is a coefficient independent to either coordinate, then $v = -\int_0^\zeta \left( \frac{\partial u}{\partial \xi} + \frac{u}{\xi} \right) d\zeta = -\frac{1+m}{2m} A\xi^{\frac{1}{m}-1}\zeta^2$. This shape function predicts that the melt surface falls in the entire melt zone. Substituting $u = A\zeta\xi^{1/m}$ into Eq. (46), and after algebraic operations, one obtains

$$A^2 g^2 \xi^{\frac{2}{m}+1} - 3PrAg\xi^{\frac{1}{m}} - \frac{6m}{m-1} \xi^2 \frac{dp}{d\xi} = 0 \qquad (47)$$

Although the coefficient $A$ can be solved locally, a potential problem is that $A$ may vary with location, which contradicts our definition of $A$. Instead, we seek a constant $A$ that makes the least square of the residuals, i.e.

$$f(A) = \int_0^\infty \left( A^2 g^2 \xi^{\frac{2}{m}+1} - 3PrAg\xi^{\frac{1}{m}} - \frac{6m}{m-1} \xi^2 \frac{dp}{d\xi} \right)^2 d\xi \qquad (48)$$

To find $A$ that makes $f(A)$ minimum, let $\frac{df(A)}{dA} = 0$, and one gets

$$W(A) = Q_1 A^3 - Q_2 A^2 + Q_3 A - Q_4 = 0 \qquad (49)$$

where

$$\begin{cases} Q_1 = \int_0^\infty 2g^4 \xi^{\frac{4}{m}+2} d\xi \\ Q_2 = 9Pr \int_0^\infty g^3 \xi^{\frac{3}{m}+1} d\xi \\ Q_3 = \int_0^\infty \left(-\frac{12m}{m-1} \xi^3 \frac{dp}{d\xi} + 9Pr^2\right) g^2 \xi^{\frac{2}{m}} d\xi \\ Q_4 = -\frac{18m}{m-1} Pr \int_0^\infty g \xi^{\frac{1}{m}+2} \frac{dp}{d\xi} d\xi \end{cases} \quad (50)$$

The Newton-Raphson method can be used for solving the coefficient $A$. Since there is no knowledge about the value of the index $m$, we take $m = 1.5, 2, 3, 4, 5$ for numerical tests. Through comparing numerical prediction with experimental data, we may find the best value of $m$.

Another possibility is that the radial velocity drops after a peak value somewhere in the melt zone. In this case, a smooth shape function $u = A\xi\zeta exp(-\xi/\xi_p)$ is assumed, where $\xi_p$ is the radius at which the radial velocity is the maximum.

$$v = -\int_0^\zeta \left(\frac{\partial u}{\partial \xi} + \frac{u}{\xi}\right) d\zeta = -\frac{A\zeta^2}{2}\left(2 - \frac{\xi}{\xi_p}\right) exp\left(-\frac{\xi}{\xi_p}\right) = -\zeta\left(\frac{1}{\xi} - \frac{1}{2\xi_p}\right) u \quad (51)$$

This shape function predicts the melt surface subsides in the domain $(0, 2\xi_p)$ whereas it rises beyond $2\xi_p$. Since there is no knowledge about $\xi_p$, we may try $\xi_p = 0.2, 0.4, 0.6, 0.8, 1.0, 1.5$ and see which value leads to the best fit to the experimental data. Substituting $u = A\xi\zeta exp(-\xi/\xi_p)$ into the momentum conservation Eq. (46) gives

$$A^2 g^2 \xi^2 - 3PrAg \, exp\left(\frac{\xi}{\xi_p}\right) - 6\xi_p \frac{dp}{d\xi} exp\left(\frac{2\xi}{\xi_p}\right) = 0 \quad (52)$$

Similar to the previous analysis, we seek a constant coefficient $A$ in terms of least square of the residuals, and we get same form of Eq. (49) with

$$\begin{cases} Q_1 = \int_0^\infty 2g^4 \xi^4 d\xi \\ Q_2 = 9Pr \int_0^\infty g^3 \xi^2 exp\left(\frac{\xi}{\xi_p}\right) d\xi \\ Q_3 = \int_0^\infty \left(9Pr^2 - 12\xi_p \xi^2 \frac{dp}{d\xi}\right) g^2 exp\left(\frac{2\xi}{\xi_p}\right) d\xi \\ Q_4 = -18Pr\xi_p \int_0^\infty g \frac{dp}{d\xi} exp\left(\frac{3\xi}{\xi_p}\right) d\xi \end{cases} \quad (53)$$

## 4. Solution of the temperature field

Now we work on the energy equation. The following temperature profile is assumed in the liquid (melt)

$$\theta_l = \left(\frac{\theta_{sat}}{g} - b_2 g\right) \zeta + b_2 \zeta^2 \quad (54)$$

which satisfies the boundary conditions in (10) and (11). The vertical gradient is

$$\frac{\partial \theta_l}{\partial \zeta} = \frac{\theta_{sat}}{g} - b_2 g + 2b_2 \zeta \tag{55}$$

At the solid-melt interface, $\frac{\partial \theta_l}{\partial \zeta}|_{\zeta=0} = \frac{\theta_{sat}}{g} - b_2 g$; at the melt-vapor interface $\frac{\partial \theta_l}{\partial \zeta}|_{\zeta=g} = \frac{\theta_{sat}}{g} + b_2 g$.

Now we solve Eq. (3b) using the assumed temperature profile and solved flow velocity. Integrating Eq. (3b) with respect to $\zeta$ from 0 to $g$, one gets

$$\frac{\partial}{\partial \xi}\int_0^g u\theta_l d\zeta + \int_0^g \theta_l \frac{u}{\xi} d\zeta + v(\xi, g)\theta_{sat} = 2b_2 g \tag{56}$$

Substituting Eq. (55) into Eq. (23), one gets

$$\{1 + \Lambda[\theta_{sat}(1-\theta_m) + \theta_m]\}exp\left[H_{lv}\left(1 - \frac{1}{\theta_{sat}(1-\theta_m)+\theta_m}\right)\right] = I'\sqrt{\theta_{sat}(1-\theta_m) + \theta_m}\left[\exp(-\xi^2) - k'\left(\frac{\theta_{sat}}{g} + b_2 g\right)\right] \tag{57}$$

Let $\theta_{sat}(1 - \theta_m) + \theta_m = \varphi$, one converts (57) into

$$(1 + \Lambda\varphi)exp\left[H_{lv}\left(1 - \frac{1}{\varphi}\right)\right] = I'\sqrt{\varphi}\left[\exp(-\xi^2) - k'\left(\frac{\varphi-\theta_m}{g(1-\theta_m)} + b_2 g\right)\right] \tag{58}$$

After algebraic operations, (58) can be rewritten as

$$A(1 + \Lambda\varphi)exp\left(-\frac{B}{\varphi}\right) + (C\varphi - D)\sqrt{\varphi} = 0 \tag{59}$$

where $A = \frac{1}{I'}exp(H_{lv})$, $B = H_{lv}$, $C = \frac{k'}{g}\frac{1}{1-\theta_m}$, $D = \exp(-\xi^2) - k'\left[b_2 g - \frac{\theta_m}{g(1-\theta_m)}\right]$.

In the pure conduction model, $b_2 = 0$, and the temperature can be solved directly through the Newton-Raphson method. In the convection model, $b_2$ is a variable and is constrained by the energy balance equation like the one expressed in (56). A convenient way is to take the value of $b_2$ obtained at the last step as an approximate to calculate $\theta_{sat}$ if the time interval is sufficiently small. Once $\theta_{sat}$ is solved in (59), the value of $b_2$ is updated in (56) by using current flow velocity and temperature.

## 5. Results and discussion

We first simulate the case of pure conduction and then conduction-flow cases, and compare them with experimental data. The thermophysical properties used are listed in Table 1. The laser intensity at the center varies from $5.3 \times 10^{10}$ W m$^{-2}$ to $14.3 \times 10^{10}$ W m$^{-2}$ in experiments of [8], and we explored the range from $2.5 \times 10^{10}$ W m$^{-2}$ to $16.0 \times 10^{10}$ W m$^{-2}$ in the numerical modeling. Similar to Zhang and Faghri [8], we ignored the effect of absorptivity of the target material. The experimental material removal rate was obtained by scaling micrographs of single shot drilled holes for pulse duration of 700 μs and laser radius of 0.254 mm at the Pratt and Whitney drilling facility at North Haven, CT. The same laser pulse duration, and the same beam intensity profile are taken in the current model as those described in Zhang and Faghri [8]. The results are shown in Fig.2. The time step used in the simulation is mainly $1.0 \times 10^{-7}$. Other time intervals like $5.0 \times 10^{-8}$, $1.0 \times 10^{-8}$ were also tested, and they led to a convergent solution. The spatial resolution was 0.0125. Running a case generally takes about 5 minutes. It can be seen that the predicted material removal rates increased with the laser beam intensity, generally consistent with the experimental data. It was also observed that the predicted results from the model without convection was very close to those predicted from models where melt flow was fully considered. The rising radial velocity models generally predicted a slightly higher removal rate than the rising-fall models. All models with melt flow being considered produced results very close to each other within 2% difference for all cases with rising profiles and rising-fall profiles.

Figure 3 shows the drilled hole profile history predicted from a rising-fall flow model where $\xi_b = 1$ and laser beam center intensity $7.5 \times 10^{10}$ [W m$^{-2}$]. It was observed that the hole developed much faster in the drill direction than in the lateral direction. These profiles are similar to each other, and are similar to the profiles predicted by Zhang and Faghri [8]. It is observed that the thickness of the melt layer increases in radial direction and the thickest melt layer occurs near the edge of the melt zone where the vaporization starts to become ignorable. It is also observed that the vertical drilling rate is almost constant.

Table 1. Thermophysical properties of test material

| | | |
|---|---|---|
| Latent heat of melting | $h_{ls}$ | $2.31 \times 10^4$ [J kg$^{-1}$] |
| Latent heat of vaporization | $h_{lv}$ | $6.444 \times 10^6$ [J kg$^{-1}$] |
| Density of melt | $\rho$ | $8.4 \times 10^3$ [kgm$^{-3}$] |
| Vapor molar mass | M | 0.076 [kg mol$^{-1}$] |
| Initial temperature | $T_i$ | 293.15 K |
| Melting temperature | $T_m$ | 1510°C |
| Standard Saturation temperature | $T_{sat0}$ | 3170°C |
| Standard Saturation pressure | $p_0$ | $1.01325 \times 10^5$ [Nm$^{-2}$] |
| Thermal conductivity of the liquid | $k_l$ | 21.70 [Wm$^{-1}$ K$^{-1}$] |
| Thermal conductivity of the solid | $k_s$ | 52.72 [Wm$^{-1}$ K$^{-1}$] |
| Specific heat of the liquid | $c_{pl}$ | 625 [J kg$^{-1}$ K$^{-1}$] |
| Specific heat of the solid | $c_{ps}$ | 380 [J kg$^{-1}$ K$^{-1}$] |
| Radius of laser beam | R | $0.254 \times 10^{-3}$ [m] |
| Prattle Number | Pr | 0.142 |

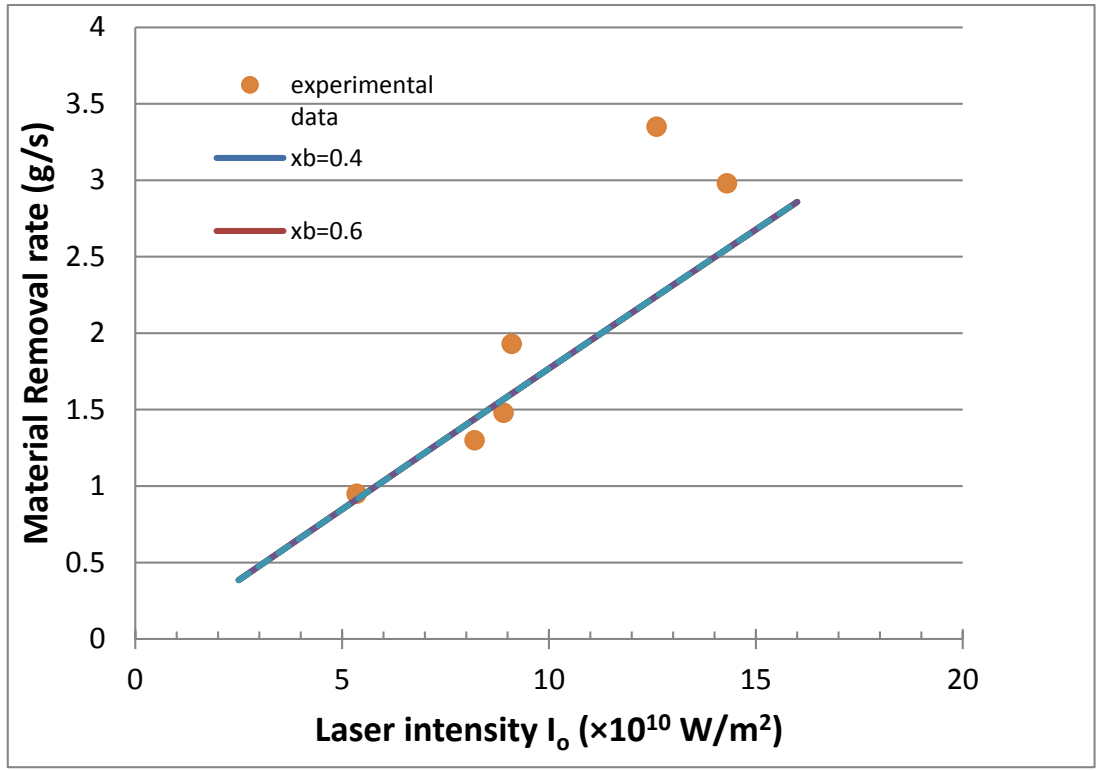

Fig. 2 Comparison of predicted and experimental material removal rates using models of rising-fall profiles

Our model is capable of predicting the temperature of the vapor or the temperature at the top of the melt. Here, we show the pressure and temperature profiles at the final stage of the drilling predicted from rising-fall models in Fig. 4 and Fig. 5. It is observed from Fig. 4 that the higher the laser beam intensity, the larger the vapor recoil pressure. The pressure inside the melt zone is not homogeneous but like Gaussian curves. It drops steeply between $\xi = 0.5\sim1.0$. Pressure at $\xi = 0.5$ is about 80% of the value at the laser beam center while it drops to one third at $\xi = 1.0$. Fig. 5 shows the temperature profiles in the corresponding models. It is observed that the temperature rises as the laser intensity increases. For all cases tested, the temperature dropped in the radial direction, but quite gradually in the majority of the melt zone, much different from pressure profile. The temperature at $\xi = 1.5$ was about 70% of the value at the center. It had a steep slope at the edge of the vapor zone.

Our modeling results also showed that the temperature and pressure are quite stable during laser beam impulse except for rapid rising at the very beginning. This pattern is confirmed in both rising profile models and rising-fall profile models (as shown in Fig. 6. and Fig.7). The peak flow velocity is predicted to experience a rapid drop at the beginning and attain a quite stable stage later from the rising models (Fig. 6a). It is slightly different from that predicted from models of rising-fall models with parameter $\xi_p < 0.7$, in which the flow gradually slows down with time (Fig. 7a). Models with parameter $\xi_p > 0.8$ generated similar results as that shown in Fig. 6. We also noticed that the solution sometimes was not available for models with a small value of $\xi_p$ (0.2 or 0.4).

The effect of the melt flow on material removal depends on many factors. One factor is the Prattle number. Fig.8 shows the melting rate, vaporization rate and melt flow velocity in a vertical direction predicted from a model of parameters list in Table 1. These rates are shown at two locations: one location is near the laser beam center $\xi = 0.0125$ and another is at $\xi = 1$. It was observed that the melting rate experienced a rapid drop as the vaporization rate rose and reached an equilibrium rather quickly. The melt flow velocity was very small, about 3-4 orders of magnitude less than the vaporization rate. Models of other profiles led to similar results. This may justify the previous model by Zhang and Faghri [8] where the flow was ignorable in the super alloy they studied. A recent study showed that natural convection does not play a significant role in melt transport and melt pool geometry [32] using a completely different approach. However, if the material is of different parameters, the melt flow might contribute to the material removal differently. To illustrate this point, we conducted a test in which all other parameters were same as that in the numerical test shown in Fig. 8 except the Prattle number. The results of the test with Prattle number 142 are shown in Fig. 9. It is observed that melt flow was able to cause about 10% of the material removal in this assumed case.

The boundary layer model was developed in this paper, but we are not presenting the results at this time. The reason is that the boundary layer assumption was found to be applicable very close to the laser beam center where the melt flows slowly. In the majority of the melt zone, the free flow zone did not exist. Another problem was that the beam center was calculated based on a pure conduction model (the radial velocity is zero but the vertical velocity is hard to estimate), while the adjacent node was based on conduction-flow which led to a differential change in the solid-melt interface. The most debatable issue probably is the boundary condition. It is unavailable at the beam center and is very roughly estimated at the second node. Even so, we still tried to evaluate the melt flow using boundary layer theory near the center and exploiting the integral method beyond the last node at which the boundary layer theory is applicable, with assumed velocity profiles following falling trend. The results were similar to that shown in previous figures. Again, the reason is because the melt flows very slowly compared with the evaporation rate.

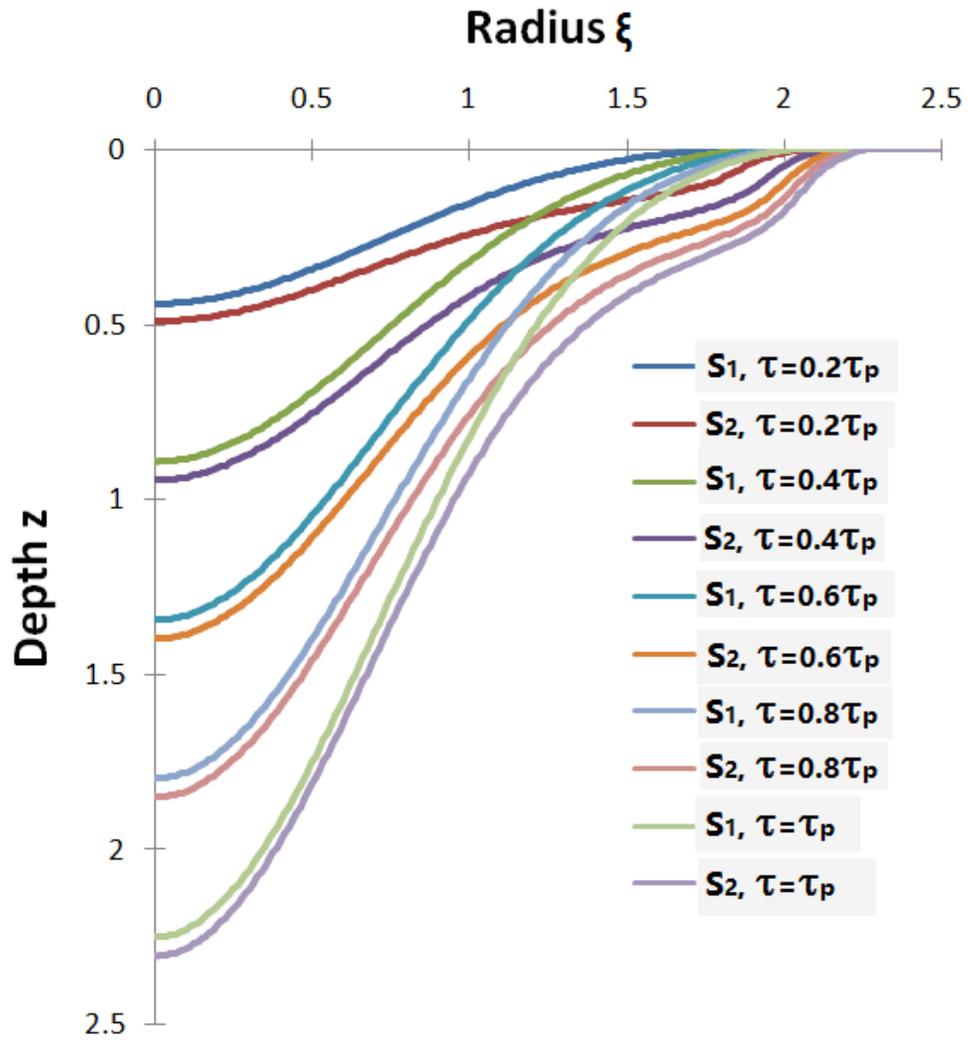
Fig. 3. Drilled hole profiles evolution predicted from a model assuming rising-fall radial velocity profile with peak velocity at $\xi_p = 1$.

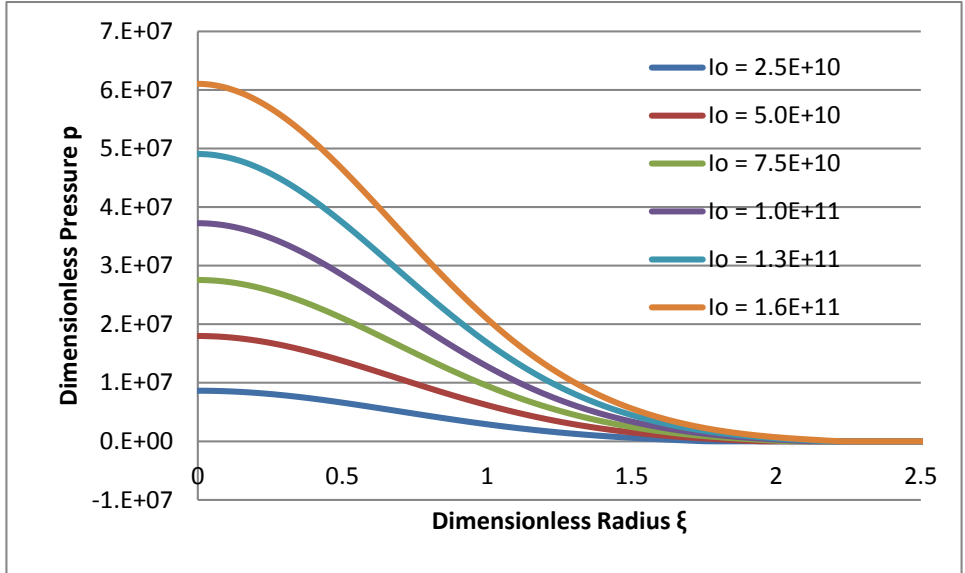

Fig. 4 Pressure profiles at $\tau = \tau_p$ with varying laser beam intensity, predicted from a model assuming rising-fall radial velocity profile with peak velocity at $\xi_p = 1$. The laser beam intensity is given at the center with unit w/m².

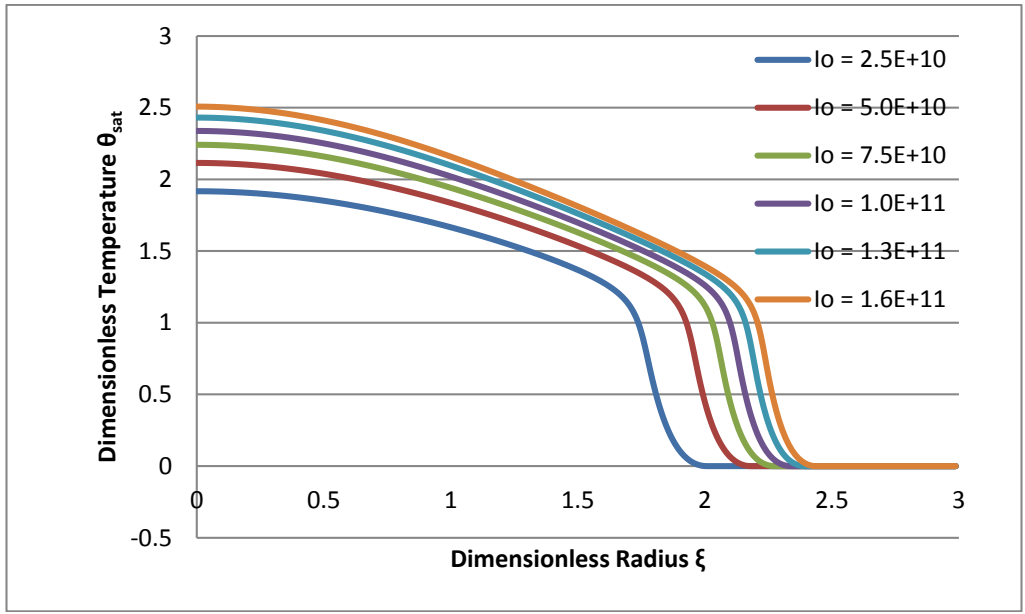

Fig. 5. Temperature profiles at $\tau = \tau_p$ with varying laser beam intensity, predicted from a model assuming a rising-fall radial velocity profile with a peak velocity at $\xi_p = 1$. The laser beam intensity is given at the center with unit w/m².

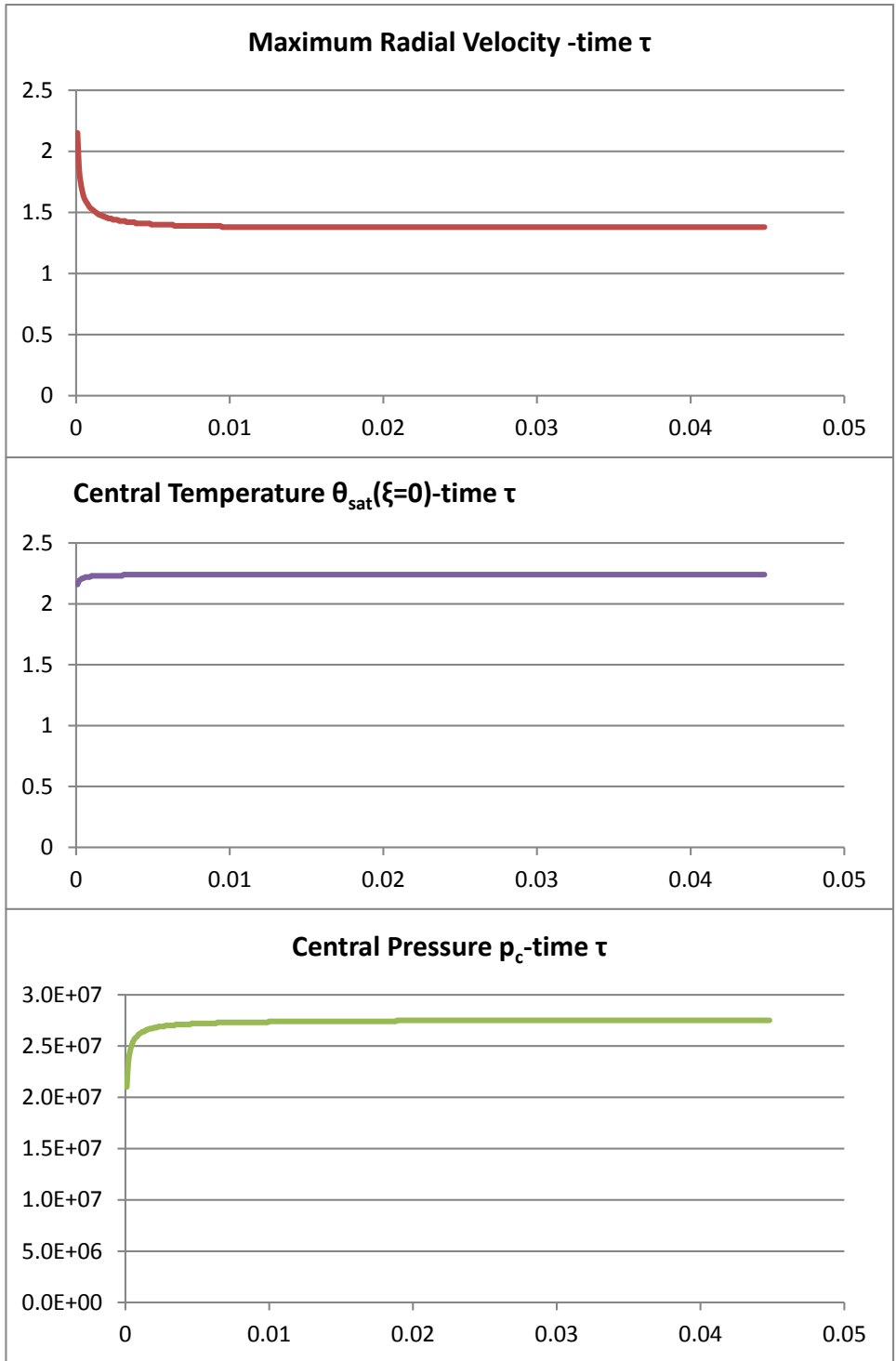

Fig. 6. Peak radial velocity, temperature and pressure history predicted from a model assuming rising radial velocity profile with index $m=2$.

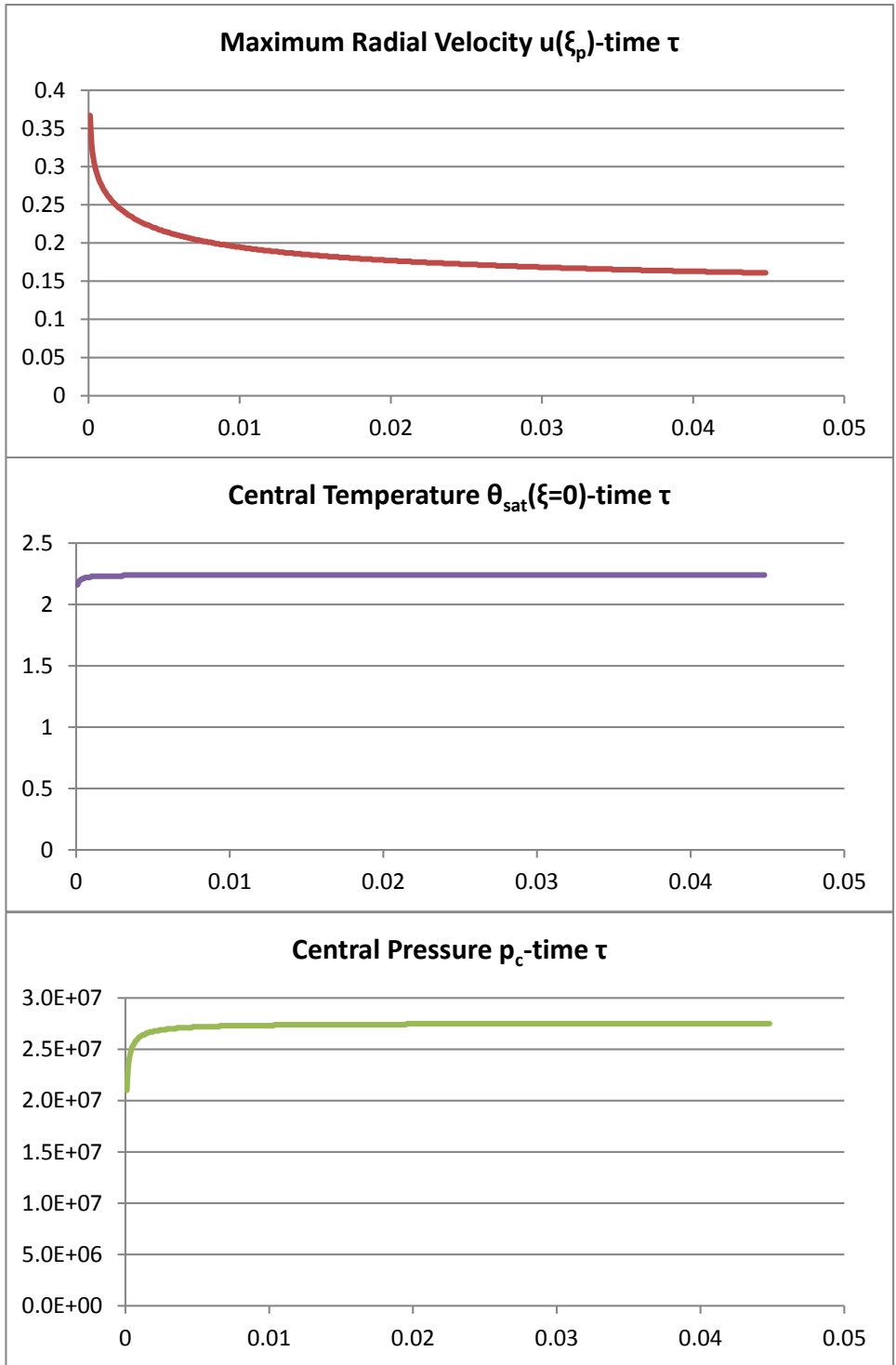

Fig. 7. Peak radial velocity, temperature and pressure history predicted from a model assuming rising-fall radial velocity profile with peak velocity at $\xi_p = 0.6$.

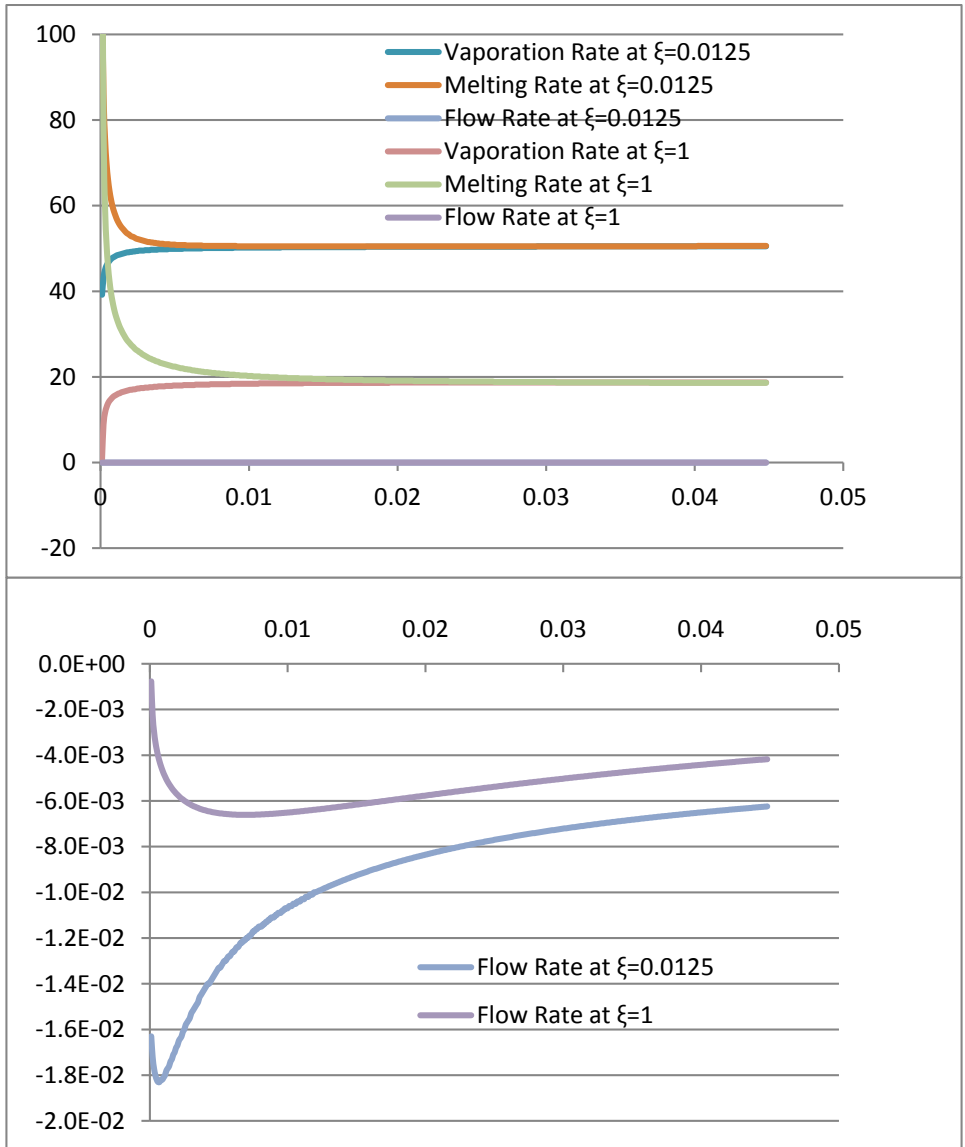

Fig. 8. (a) Melting rate, vaporization rate and the effect of the melt flow on interfacechange near the laser beam center at $\xi = 0.0125$ and at $\xi = 1$.
(b) The change of flow velocity vertical component at both locations.

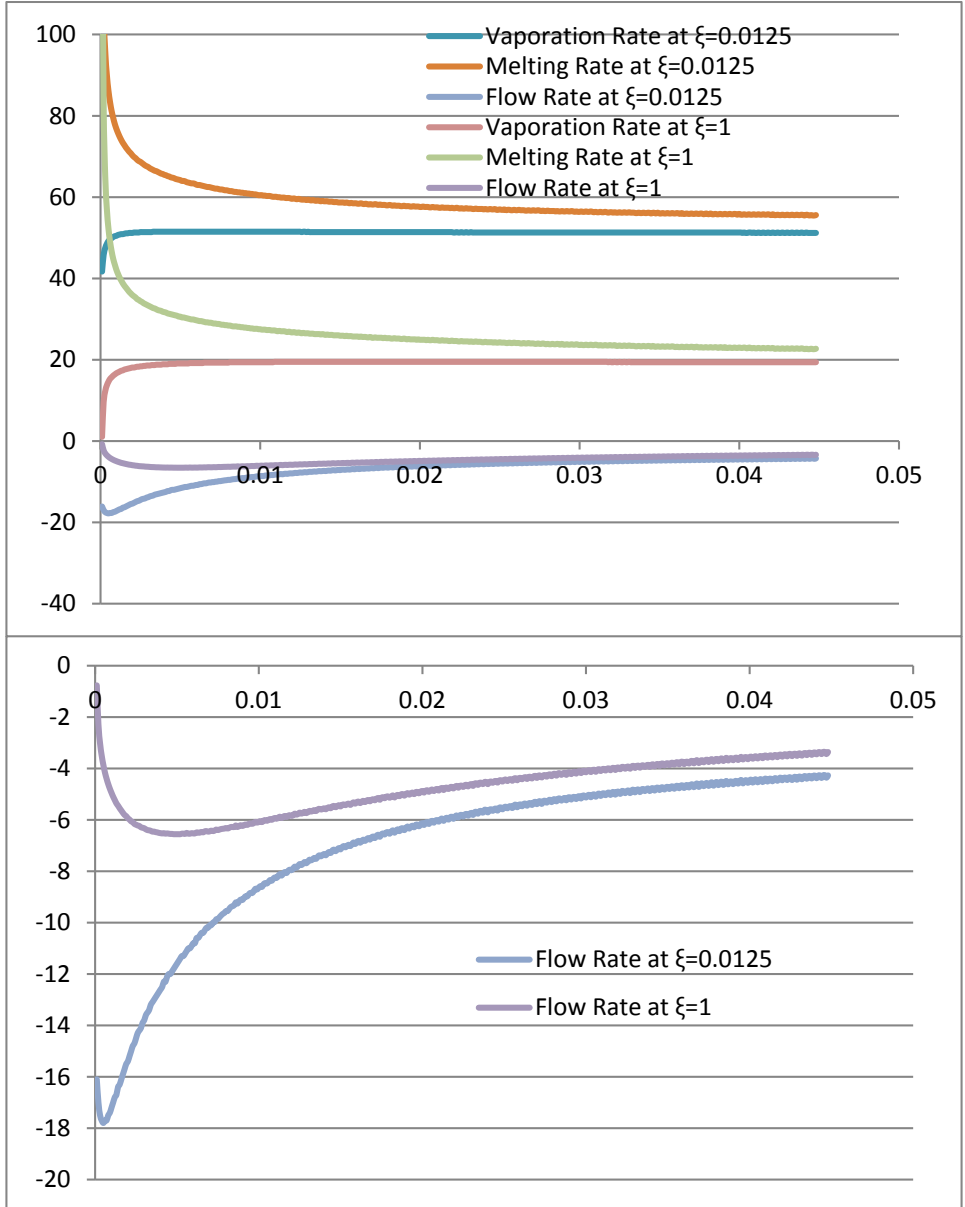

Fig. 9. (a) Melting rate, vaporization rate and the effect of the melt flow on interface change near the laser beam center $\xi = 0.0125$ and at $\xi = 1$.
(b) The change of flow velocity vertical component at both locations, similar to Fig. 8 but with a Prattle number 1000 times larger.

We attempted to find the flow pattern by comparing experimental data with models of various radial velocity profiles. This mission is not completed in the cases of very slow flow as in the case of the super alloy studied here. Later experiments may provide better data to assist with flow pattern identification.

# 6. Concluding remarks

The novelty of this investigation is that the melt flow effect on laser drilling is explicitly included in the equation of momentum conservation. By applying the non-slip boundary condition at the solid-liquid interface, the solution of the melt flow is obtained by using the boundary layer theory and integral method. An integral solution for the temperature field is also obtained. In addition, the exact solution for heat conduction is developed. The solution of Stefan problem is the principal component in the exact solution, and can be used as an approximate solution for better computational efficiency. The dependence of saturation temperature on the vapor pressure is taken into account by using the Clausius-Clapeyron equation. As compared with previous models, the proposed model is more representative of the real physics involved. Applying the new model to a super alloy, we have demonstrated that the melt flow effect could be ignored in some cases. The solutions obtained here will be further applied to more general cases to evaluate the effects of melt flow and vaporization on the laser drilling profile evolution, and to study the solid material removal efficiency.

**Acknowledgments**

This work was supported in part by the U. S. Defense Threat Reduction Agency under grant number HDTRA1-10-1-0022, and by the National Science Foundation of China under Grant number 11232003. The authors appreciate the anonymous reviewers for insightful comments that improved the paper.

## Appendix

The exact solution for time period after melting starts consists of three parts:

$$\theta_s(\xi, n, \tau) = \theta_{s1} + \theta_{s2} + \theta_{s3}, \quad \tau > \tau_m \tag{A1}$$

where

$$\theta_{s1}(n, \tau) = \theta_i\, erf\left[\frac{n}{2\sqrt{N_\alpha(\tau + \tau_0 - \tau_m)}}\right], \quad \tau > \tau_m \tag{A2}$$

in which

$$\tau_0 = \frac{\left[\frac{\theta_i k\prime}{k} exp(\xi^2)\right]^2}{4 N_\alpha} = \frac{\tau_m}{4\pi} \tag{A3}$$

Equation (A2) is the solution of the Stefan problem. The idea of adding $\tau_0$ is to avoid singularity at $\tau = \tau_m$ and to make the temperature gradient (heat flux) continuous at $\tau = \tau_m$, which also has a generated temperature profile

$$\theta_{s1}(n, \tau_m) = \theta_i\, erf\left(\frac{n}{2\sqrt{N_\alpha \tau_0}}\right) \tag{A4}$$

This profile is called "generated" because the initial condition $\theta_i$ is assumed at the moment $\tau = \tau_m - \tau_0$ for the Stefan solution. Because the temperature field at melting starting time is not the profile expressed in Eq. (A4) but the established profile shown in Eq. (29), Eq. (A2) is not the exact solution of the problem concerned. The difference between Eqs. (29) and (A4) must be corrected. For a semi-infinity object with no heat source and zero initial and boundary conditions, the temperature evolution due to an existing non-zero temperature profile $G(n)$ at a given moment can be calculated using following formula [33]

$$\theta_{s2}(n,\tau) = \frac{1}{2\sqrt{\pi N_\alpha(\tau-\tau_m)}} \int_0^\infty G(y) e^{-\frac{(n-y)^2}{4N_\alpha(\tau-\tau_m)}} dy, \tau > \tau_m \tag{A5}$$

For our problem $G(n) = \theta_s(\xi, n, \tau_m) - \theta_{s1}(n, \tau_m)$. Obviously this solution will lead to a non-zero boundary temperature at $n = 0$.

$$\theta_{s2b}(0,\tau) = \frac{1}{2\sqrt{\pi N_\alpha(\tau-\tau_m)}} \int_0^\infty G(y) e^{-\frac{y^2}{4N_\alpha(\tau-\tau_m)}} dy, \tau > \tau_m \tag{A6}$$

In order to keep the constant zero temperature at the boundary (Stefan condition), we need to add a new correction by considering the effect of boundary temperature $\theta_{s3b}(0,\tau) = -\theta_{s2b}(0,\tau)$. The solution of this Dirichlet condition is classic, and the uniqueness of the solution is proven in [33]. It can be solved by using numerical methods like finite element methods or finite difference methods. An easy way is to use pdepe function in MATLAB. The solution is now marked as $\theta_{s3}$.

Both $\theta_{s2}$ and $\theta_{s3}$ declines very fast with time, and they become ignorable as $\tau > 5\tau_m$. The Stefan solution is the major contributor. To reduce computational expense, an approximated solution is obtained by simply shifting the Stefan solution backward by $3\tau_0$, which leads to ignorable change for $\tau > 5\tau_m$ but attains significant improvement on the temperature gradient for $\tau < \tau_m$.

$$\theta_s(n,\tau) \approx \theta_i erf\left[\frac{n}{2\sqrt{N_\alpha(\tau+3\tau_0-\tau_m)}}\right] \tag{A7}$$

$$\frac{\partial \theta_s}{\partial n}(0,\tau) \approx \frac{\theta_i}{\sqrt{\pi N_\alpha \tau_m}\sqrt{\frac{\tau}{\tau_m}-1+\frac{3}{4\pi}}} \tag{A8}$$

Since the coordinate origin moves at the rate of $V_n = \frac{\partial S_2}{\partial \tau} / \sqrt{1+\left(\frac{\partial S_2}{\partial \xi}\right)^2}$, the effective temperature gradient at a static coordinate system will be $\frac{\partial \theta_s}{\partial n}(0,\tau) + \theta_i \frac{V_n}{N_\alpha}$ in terms of same energy fluency.

The curvature-corrected temperature gradient is $C_c\left[\frac{\partial \theta_s}{\partial n}(0,\tau) + \theta_i \frac{V_n}{N_\alpha}\right]$ where

$$C_c = \begin{cases} \frac{ad}{2sin^{-1}\frac{ad}{2}} & a > 0 \\ 1 & a = 0 \\ \frac{2sin^{-1}\frac{ad}{2}}{ad} & a < 0 \end{cases} \tag{A9}$$

and

$$a = -\frac{\partial^2 S_2}{\partial \xi^2}\left[1 + (\frac{\partial S_2}{\partial \xi})^2\right]^{-3/2} \tag{A10}$$

is the curvature of the melting interface. The minus sign is due to the $S_2$ point to $z$ direction. In numerical tests, $d = \Delta\xi\sqrt{1 + \left(\frac{\partial S_2}{\partial \xi}\right)^2}$, where $\Delta\xi$ is the interval of $\xi$. The interval can be taken as small enough to make $ad < 2$.